\documentclass [reprint,aps,amsmath,amssymb,twoside,prb,showkeys,superscriptaddress]{revtex4-1}
\usepackage{verbatim}
\usepackage{graphicx}
\usepackage[T1]{fontenc}
\usepackage{textcomp}
\usepackage{txfonts}
\usepackage{tensor}
\usepackage{cancel}
\usepackage{color}
\newcommand{\HCd}{\mathcal{H}}

\def\HCdt0{\tilde{\HCd}_{0}}
\newcommand{\afffias}{Frankfurt Institute for Advanced Studies (FIAS), Ruth-Moufang-Strasse~1, 60438 Frankfurt am Main, Germany}
\newcommand{\affjwg}{Goethe-Universit\"at, Max-von-Laue-Strasse~1, 60438~Frankfurt am Main, Germany}
\newcommand{\affbgu}{Physics Department, Ben-Gurion University of the Negev, Beer-Sheva 84105, Israel}
\newcommand{\affbahamas}{Bahamas Advanced Study Institute and Conferences, 4A Ocean Heights, Hill View Circle, Stella Maris, Long Island, The Bahamas}
\begin{document}

\preprint{}

\title{A correspondence between $1^{st}$ and $2^{nd}$ order formalism by a metricity constraint}

\author{David Benisty}
\email{benidav@post.bgu.ac.il}
\affiliation{\afffias}\affiliation{\affjwg}\affiliation{\affbgu}
\author{Eduardo I. Guendelman}
\email{guendel@bgu.ac.il}
\affiliation{\afffias}\affiliation{\affbgu}\affiliation{\affbahamas}

\date{\today}

\begin{abstract}
A way to obtain a correspondence between the first order and second order formalism is studied. By introducing a Lagrange multiplier coupled to the covariant derivative of the metric, a metricity constraint is implemented. The new contributions which comes from the variation of the Lagrange multiplier  transforms the field equations from the first order to the second order formalism, yet the action is formulated in the first order.
In this way all the higher derivatives terms in the second order formalism appear as derivatives of the Lagrange multiplier. Using the same method for breaking metricity condition and building conformal invariant theory is briefly discussed,
so the method goes beyond just the study of first order or second formulations of gravity, in fact vast new possible theories of gravity are envisioned this way.     
\begin{description}
\item[Key Words]
First order, second order formalism, metricity, conformal symmetry
\end{description}
\end{abstract}

\pacs{Valid PACS appear here}
\maketitle
\section{introduction}
There are two main formulations that are used in gravity theories. The $\textbf{first order formalism}$, which is also called the Palatini formalism \cite{Palatini} treats  the metric and the connection as independent degrees of freedom. The connection is obtained through the solution of the equations of motion. In general, the solution does not result in the connection being the Levi Civita or Christoffel symbol. In contrast to that, in the $\textbf{second order formalism}$ the connection is assumed to be the Levi Civita or Christoffel symbol:
\begin{equation}
\{ ^\rho_{\mu \nu} \} = \frac{1}{2} g^{\rho\lambda} (g_{\lambda\mu,\nu}+g_{\lambda\nu,\mu}-g_{\mu\nu,\lambda})
\end{equation}
and appears in the action in this way, not by being an independent degree of freedom.
Those two formulations are used independently by different researchers and in general those two are inequivalent formulation for similarly looking gravity theories in terms of the dependence on the scalar curvature tensors and scalars. Only for Lovelock theories \cite{Lovelock}, which includes Einstein Hilbert action at the first order, both formulations will yield the same equations of motion and the connection will be in both cases the Christoffel symbol \cite{Exirifard:2007da}. 

Many of the modified theories of gravity that we consider, as $f(R)$ gravity and higher curvature terms $f(R,R_{\mu\nu}R^{\mu\nu}, R_{\alpha\beta\gamma\delta}R^{\alpha\beta\gamma\delta})$, are viable and can exist in a wide parameter range. To just a few  examples where using  higher curvature terms has been done are in inflationary models, first of all in the second order formalism in the Starobinski model $R^2$, a higher curvature terms in context of inflationary solutions \cite{Gorbunov:2013dqa}\cite{Myrzakulov:2014hca}\cite{Bamba:2014jia}\cite{Sebastiani:2013eqa} or quadratic Gauss Bonnet inflation \cite{DeLaurentis:2015fea}. In the Palatini formalism or in the metric formalism as well, $f(R)$ theories of gravity are used to describe the accelerated cosmological expansion \cite{Sotiriou:2008rp}. Those theories are consistent with the observational constraint for a range of parameters of the theory. The subject of alternative theories of gravity has been very active in order to provide a new approach to the puzzles of cosmology, like the Dark Matter and Dark Energy questions and for other fields as  black holes and neutron stars structure and merger. Each alternative theory of gravity has then to be compared with observational data, etc.

While for general relativity and other Lovelock theories, the first order and the second order formalisms give just two different variational presentations of the theory and the name "formalism" is indeed justified, for more generic Lagrangian, they are not , 
since a similar looking Lagrangian, in terms of its dependence on the curvature tensors, etc., leads to a different theory in the first order formalism and in the second order formalism. The name "formalism" is therefore somewhat misleading in this case, although it has continued to be used anyway, but instead of just being only a formalism, it represents a way to build a different theory of gravity from the same looking Lagrangian.
\section{Basic formulation}
We are not going to deal with a specific theory of gravity. Instead, the objective of this paper is to show that  all known theories of gravity (and some more that could be formulated as we will see) and in particular the second order formulation of a gravitational theory can be formulated in a first order form. Indeed we will see that in a first order formulation and by using also a Lagrange multiplier tensor field $k^{\alpha\beta\gamma}$ which couples to the covariant derivative of the metric, the equation of the Lagrange multiplier that enforces the vanishing of the covariant derivative of the metric (metricity condition) we can convert the first order equations of motion to reproduce the the equations of motion in the second order formulation, however the action is still formulated in the first order formalism:
\begin{equation}\label{t}
\mathcal{L}(g)\,_{2^\textbf{order}} \Leftrightarrow \mathcal{L}(g,\Gamma) + k^{\alpha\beta\gamma}  g_{\alpha\beta;\gamma} \,_{1^\textbf{order}}
\end{equation}
The variation with respect to $k^{\alpha\beta\gamma}$ gives the metricity condition:
\begin{equation}\label{met}
g_{\alpha\beta;\gamma} = 0 \quad \Rightarrow \quad \Gamma^{\rho}_{\mu\nu} = \{ ^\rho_{\mu \nu} \}
\end{equation}
This type of Lagrange multiplier was first considered for the purpose of giving a canonical conjugate momentum to the metric in the context of a covariant  gauge theory of gravity \cite{Struckmeier:2017vkf}\cite{Struckmeier:2017stl}, but considered only for a very special case in a Hamiltonian approach. An equivalent constraint was taken into account in \cite{Hehl1}\cite{Hehl:1976my}\cite{Hehl:1981ed} which has a similar conclusions, but was not discussed with higher curvature terms. For other different constraints, the vanishing of the covariant derivative of the metric can be explored giving a new possibilities as we will discuss in the case of a formulation for conformal symmetry, but also many other possibilities could be considered, leading to the possibility of formulating many new theories of gravity. 
\section{The connection between the variations}
For the additional term in the action that introduces a Lagrange multiplier:
\begin{equation}
\mathcal{S}_{(\kappa)} = \int d^4x \, \sqrt{-g}k^{\alpha\beta\gamma}  g_{\alpha\beta;\gamma}
\end{equation}
we consider a 3 index tensor. The variation with respect to this tensor gives the metricity condition, which cause the connection to be the Christoffel symbol, even if the action is formulated in first order formalism. As we will see this formulation gives the same equations of motion as the second order formalism, but the equations appear as up to second order differential equations, even for higher curvature action. The variation with respect to the connection gives the tensors:
\begin{equation}\label{g}
\frac{\delta \mathcal{L}(\kappa)}{\delta \Gamma^{\rho}_{\mu\nu}} = - k^{\alpha\mu\nu}g_{\rho\alpha} - k^{\alpha\nu\mu}g_{\rho\alpha}
\end{equation}
with a symmetrization between the components $\mu$ and $\nu$. 

The variation with respect to the metric is:
\begin{equation}\label{eomm}
G^{\mu\nu}_{(\kappa)} = \frac{\delta \mathcal{L}(\kappa)}{\delta g_{\mu\nu}} = -k^{\mu\nu\lambda}_{;\lambda}
\end{equation}
Because of the new contribution to the field equation $G^{\mu\nu}_{(\kappa)}$ the complete field equation will contain additional terms which make the first order field equations to be equivalent to the field equation under second order formalism. 

For obtain the contribution for the variation with respect to the metric, we have to use the variation with respect to the $\Gamma$. In that way, the value of the tensor $ k^{\alpha\beta\gamma}$ will appear: 
\begin{equation}\label{eq1}
g^{\rho\sigma}\frac{\partial \mathcal{L}(\kappa)}{\partial \Gamma^{\rho}_{\mu\nu}} = -k^{\sigma\mu\nu} - k^{\sigma\nu\mu}
\end{equation}
By changing the indices we get the relations:
\begin{equation}\label{eq2}
g^{\rho\nu}\frac{\partial \mathcal{L}(\kappa)}{\partial \Gamma^{\rho}_{\mu\sigma}} = -k^{\nu\mu\sigma} - k^{\nu\sigma\mu}
\end{equation}
\begin{equation}\label{eq3}
g^{\rho\mu}\frac{\partial \mathcal{L}(\kappa)}{\partial \Gamma^{\rho}_{\nu\sigma}} = -k^{\mu\nu\sigma} - k^{\mu\sigma\nu}
\end{equation}
Adding Eq. (\ref{eq1}) into Eq. (\ref{eq2}) minus Eq. (\ref{eq3}) gives:
\begin{equation}\label{eq4}
g^{\rho\sigma}\frac{\partial \mathcal{L}(\kappa)}{\partial \Gamma^{\rho}_{\mu\nu}}+g^{\rho\nu}\frac{\partial \mathcal{L}(\kappa)}{\partial \Gamma^{\rho}_{\mu\sigma}}-g^{\rho\mu}\frac{\partial \mathcal{L}(\kappa)}{\partial \Gamma^{\rho}_{\nu\sigma}}=-2k^{\nu\sigma\mu}
\end{equation}

which is the value of the tensor, without any symmetrization of the indices. Therefore the variation with respect to the metric, which comes from the term $-\frac{\delta \mathcal{L}(\kappa)}{\delta g_{\sigma\nu}} =k^{\nu\sigma\mu}_{;\mu}$ will given by a derivative of the tensor from equation (\ref{eq4}): 
\begin{equation}
\frac{\delta \mathcal{L}(\kappa)}{\delta g_{\sigma\nu}}  = \frac{1}{2}\nabla_{\mu}(g^{\rho\sigma}\frac{\partial \mathcal{L}(\kappa)}{\partial \Gamma^{\rho}_{\mu\nu}}+g^{\rho\nu}\frac{\partial \mathcal{L}(\kappa)}{\partial \Gamma^{\rho}_{\mu\sigma}}-g^{\rho\mu}\frac{\partial \mathcal{L}(\kappa)}{\partial \Gamma^{\rho}_{\nu\sigma}})
\end{equation}
as we see in Eq. (\ref{connection}). Indeed solving the tensor $k^{\mu\nu\lambda}$ and inserting back into Eq. (\ref{eomm}) gives:  
\begin{equation}\label{connection}
\frac{\delta \mathcal{L}(\kappa)}{\delta g_{\sigma\nu}}  = \frac{1}{2}\nabla_{\mu}(g^{\rho\sigma}\frac{\partial \mathcal{L}(\kappa)}{\partial \Gamma^{\rho}_{\mu\nu}}+g^{\rho\nu}\frac{\partial \mathcal{L}(\kappa)}{\partial \Gamma^{\rho}_{\mu\sigma}}-g^{\rho\mu}\frac{\partial \mathcal{L}(\kappa)}{\partial \Gamma^{\rho}_{\nu\sigma}})
\end{equation}
where the terms in the right hand side represents the additional terms that appear in the second order formalism. One option for obtain the contributions into the field equation is to solve $k^{\alpha\beta\gamma}$. The direct way is by using this equation, that gives the new contributions for the second order formalism into the field equation, from the variation with respect to the connection $\Gamma^{\rho}_{\mu\nu}$. Let's see a simple example for the correspondence. 

\section{A Higher curvature terms example}
To see how this idea is implemented, let's take a form of action up to second power for curvature terms \cite{Borunda:2008kf}\cite{Olmo:2013lta}\cite{Barrientos:2018cnx}:
\begin{equation}
\mathcal{L}(g)\,_{2^\textbf{order}} = R + \frac{\alpha}{2} R^2 +\frac{\beta}{2} R_{\mu\nu}R^{\mu\nu} +\frac{\gamma}{2} R_{\alpha\beta\gamma\delta} R^{\alpha\beta\gamma\delta} 
\end{equation}
The variation of the action with respect to the metric in the $\textbf{second order formalism}$ gives the terms:
\begin{subequations}\label{f2o}
\begin{equation}
G^{\mu\nu}_{(0)} = R^{\mu\nu} - \frac{1}{2}g^{\mu\nu}R
\end{equation}
\begin{equation}
G^{\mu\nu}_{(\alpha)} = R (R^{\mu\nu} - \frac{1}{4}g^{\mu\nu}R) - \nabla^{\mu}\nabla^{\nu} R +g^{\mu\nu} \Box R
\end{equation}
\begin{equation}
\begin{split}
G^{\mu\nu}_{(\beta)} = R^{\mu\gamma}R^{\nu}_{\gamma} - \frac{1}{4} g^{\mu} R^{\alpha\beta}R_{\alpha\beta} \\
-\frac{1}{2}\nabla_{\gamma}(\nabla^{\mu}R^{\nu\gamma}+\nabla^{\nu}R^{\mu\gamma})+\frac{1}{2}\Box R^{\mu\nu}+\frac{1}{4} \Box R 
\end{split}
\end{equation}
\begin{equation}
\begin{split}
G^{\mu\nu}_{(\gamma)} = R^{\mu\alpha\beta\gamma} R^{\nu}_{\alpha\beta\gamma} - \frac{1}{4} g^{\mu\nu} R^{\alpha\beta\gamma\delta} R_{\alpha\beta\gamma\delta} +\\  (\nabla_{\alpha} \nabla_{\beta} + \nabla_{\beta} \nabla_{\alpha}) R^{\mu\alpha\nu\beta}
\end{split}
\end{equation}
\end{subequations}
Where the complete variation is the sum of the partial ones:
\begin{equation}
G^{\mu\nu}=G^{\mu\nu}_{(0)} + \alpha G^{\mu\nu}_{(\alpha)} + \beta G^{\mu\nu}_{(\beta)} + \gamma G^{\mu\nu}_{(\gamma)}
\end{equation}
In the vacuum case (no matter) $G^{\mu\nu} = 0$.
These equations of motion in second order formalism should coincide to the equations of motion of the action 
with the Lagrange multiplier in first order formalism as (\ref{t})
The variation with respect to the Lagrange multiplier $k^{\alpha\beta\gamma}$ forcing metricity condition (\ref{met}) and the connection being Christoffel symbol. The variation with respect to the connection gives:
\begin{subequations}\label{vg}
\begin{equation}
K_{\lambda \, (0)}^{\mu\nu}  = \nabla_\rho (g^{\mu\rho}\delta^{\nu}_{\lambda}-g^{\nu\mu}\delta^{\rho}_{\lambda}) 
\end{equation}
\begin{equation}
K_{\lambda \, (\alpha)}^{\mu\nu} =  (g^{\mu\nu} \nabla_{\lambda} -\frac{1}{2}\delta^{\nu}_{\lambda} \nabla^{\mu}-\frac{1}{2} \delta^{\mu}_{\lambda} \nabla^{\nu})R
\end{equation}
\begin{equation}
K_{\lambda \, (\beta)}^{\mu\nu} =  \nabla_{\lambda}R^{\mu\nu} - \frac{1}{4}\delta^{\mu}_{\lambda} \nabla^{\nu}R -\frac{1}{4}\delta^{\nu}_{\lambda} \nabla^{\mu}R
\end{equation}
\begin{equation}
K_{\lambda \, (\beta)}^{\mu\nu} = \nabla_{\sigma} R_{\lambda}^{\,\mu\sigma\nu} + \nabla_{\sigma} R_{\lambda}^{\,\nu\sigma\mu}
\end{equation}
\end{subequations}
Where also here the complete variation is the sum of the partial ones: 
\begin{equation}
- k^{\mu\beta\nu} g_{\lambda\beta} - k^{\nu\beta\mu} g_{\lambda\beta}+K^{\mu\nu}_{\lambda \, (0)} + \alpha K^{\mu\nu}_{\lambda \, (\alpha)} + \beta K^{\mu\nu}_{\lambda \, (\beta)} +\gamma K^{\mu\nu}_{\lambda \, (\gamma)} = 0
\end{equation}
Because of the metricity condition (\ref{met}), the variation of $R$ with respect to the connection gives identically zero
$K^{\mu\nu}_{\lambda \, (0)} = 0$. Therefore we get:  
\begin{equation}\label{kvar}
k^{\mu\beta\nu} g_{\lambda\beta} + k^{\nu\beta\mu} g_{\lambda\beta} = \alpha K^{\mu\nu}_{\lambda \, (\alpha)} + \beta K^{\mu\nu}_{\lambda \, (\beta)} +\gamma K^{\mu\nu}_{\lambda \, (\gamma)}
\end{equation}
The field equation are obtained from the variation with respect to the metric (in the $\textbf{first order formalism}$):
\begin{subequations}\label{f1o}
\begin{equation}
G^{\mu\nu}_{(\kappa)} = - k^{\mu\nu\lambda}_{;\lambda}
\end{equation}
\begin{equation}
G^{\mu\nu}_{(0)} = R^{\mu\nu} - \frac{1}{2}g^{\mu\nu}R
\end{equation}
\begin{equation}
G^{\mu\nu}_{(\alpha)} = R (R^{\mu\nu} - \frac{1}{4}g^{\mu\nu}R)
\end{equation}
\begin{equation}
G^{\mu\nu}_{(\beta)} = R^{\mu\gamma}R^{\nu}_{\gamma} - \frac{1}{4} g^{\mu} R^{\alpha\beta}R_{\alpha\beta}
\end{equation}
\begin{equation}
G^{\mu\nu}_{(\gamma)} = R^{\mu\alpha\beta\gamma} R^{\nu}_{\alpha\beta\gamma} - \frac{1}{4} g^{\mu\nu} R^{\alpha\beta\gamma\delta} R_{\alpha\beta\gamma\delta}
\end{equation}
\end{subequations}
with the complete field equation:
\begin{equation}
G^{\mu\nu}=G^{\mu\nu}_{(\kappa)}+G^{\mu\nu}_{(0)} + \alpha G^{\mu\nu}_{(\alpha)} + \beta G^{\mu\nu}_{(\beta)} +\gamma G^{\mu\nu}_{(\gamma)} 
\end{equation}

From $G^{\mu\nu}_{(\kappa)}$ we get the contribution to to field equation which transforms the equations of motion from the original terms of the first order formalism into the additional term in the second order 
formalism. To show that, let's use Eq. (\ref{connection}) by substituting all of the $K^{\mu\nu}_\lambda=\frac{\delta\mathcal{L}}{\delta \Gamma^{\mu\nu}_{\lambda}}$ terms from Eq. (\ref{vg}). The contribution from $G^{\mu\nu}_{(\kappa)}=-k^{\mu\nu\lambda}_{;\lambda}$ gives:
\begin{subequations}
\begin{equation}
G^{\mu\nu}_{(\alpha)}\,_{2^\textbf{order}}=G^{\mu\nu}_{(\alpha)}\,_{1^\textbf{order}} + \alpha\left[ - \nabla^{\mu}\nabla^{\nu} R +g^{\mu\nu} \Box R \right]
\end{equation}
\begin{equation}
\begin{split}
G^{\mu\nu}_{(\beta)}\,_{2^\textbf{order}}=G^{\mu\nu}_{(\beta)}\,_{1^\textbf{order}}+\\\beta\left[-\frac{1}{2}\nabla_{\gamma}(\nabla^{\mu}R^{\nu\gamma}+\nabla^{\nu}R^{\mu\gamma})+\frac{1}{2}\Box R^{\mu\nu}+\frac{1}{4} \Box R \right]
\end{split}
\end{equation}
\begin{equation}
G^{\mu\nu}_{(\gamma)}\,_{2^\textbf{order}}=G^{\mu\nu}_{(\gamma)}\,_{1^\textbf{order}}+\gamma\left[(\nabla_{\alpha} \nabla_{\beta} + \nabla_{\beta} \nabla_{\alpha}) R^{\mu\alpha\nu\beta}\right]
\end{equation}
\end{subequations}
which are the missing terms that shifted the field equation from the original first order field equation (\ref{f2o}) into the field equation in second order formalism (\ref{f1o}), using Bianchi identities and symmetrization of the indices.
From the variation of the correspondence (\ref{t}) with respect to the metric we obtain that in general 
\begin{equation}
G^{\mu\nu}_{(\gamma)}\,_{2^\textbf{order}}=G^{\mu\nu}_{(\gamma)}\,_{1^\textbf{order}} + G^{\mu\nu}_{(\kappa)}
\end{equation}
as shown for the example above.
\section{The Path integral approach}
A formal argument valid even in the quantum case can be formulated in the path integral approach of \cite{pi}. Consider the path integral over all the field variables ($g,k,\Gamma$) independently of each other, the path integral is:
\begin{equation}
\mathcal{Z} = \int \mathcal{D}k \mathcal{D}g \mathcal{D}\Gamma e^{i\int d^4x \sqrt{-g}(\mathcal{L}(g,\Gamma) + k^{\alpha\beta\gamma} g_{\alpha\beta;\gamma})}
\end{equation}
Performing the integral over the field $k$ we obtain the delta function enforcing the metricity relation:  
\begin{equation*}
=\int \delta(g_{\alpha\beta;\gamma}) \mathcal{D}g \mathcal{D}\Gamma e^{i\int d^4x \sqrt{-g}\mathcal{L}(g,\Gamma)}
\end{equation*}
Since we know that the metricity condition enforces the connection to be equal to the Christoffel symbol (\ref{met}): 
\begin{equation*}
\sim \int \delta(\Gamma^{\rho}_{\mu\nu} - \{ ^\rho_{\mu \nu} \}) \mathcal{D}g \mathcal{D}\Gamma e^{i\int d^4x \sqrt{-g}\mathcal{L}(g,\Gamma)}
\end{equation*}
therefore after integration over $\Gamma$ we obtain the path integral in the second order formulation:
\begin{equation}\label{fPi}
\mathcal{Z} = \int \mathcal{D}g \, e^{i\int d^4x \sqrt{-g}\mathcal{L}(g)}
\end{equation}
where the $\Gamma$ has been replaced by the Christoffel symbol every where. The final path integral (\ref{fPi}) is the path integral which represents an action under the second order formalism. This argument shows us that the second order formalism, which contains higher derivatives in the action, can be put in the first order form without higher derivatives. This fact should be useful for some semi-quantum version of gravity theories with higher curvatures terms. In front of quantizing the action in the second order formalism with higher derivatives, one could use the first order formalism with the metricity constraint, which formulate the action in lower derivatives of the metric and connection independently. 

Concerning the initial value problem, since the models considered at the end are equivalent to a theory formulated in the second order formalism, the formulation of the initial value problem is also equivalent, nevertheless, here, with the new variables introduced here, including the Lagrange multiplier field, all the initial conditions can be expressed in terms as initial values for the fields (including for the Lagrange multiplier field). This is similar to the Hamiltonian formalism where the canonically conjugate variables are introduced and the initial condition involve the initial values of the fields and their canonically conjugate momenta.

\section{A conformal invariant case}
A generalized constraint on the metric that respects conformal invariance \cite{weyl}\cite{W}\cite{Romero:2012hs} could be used from those notions. By introducing a vector field $A_\mu$ into the constraint:
\begin{equation}
\tilde{\mathcal{L}}(\kappa)=\sqrt{-g}k^{\alpha\beta\gamma}  (g_{\alpha\beta;\gamma}- e g_{\alpha\beta}A_{\gamma})
\end{equation}
a conformal symmetry emerges. Where $e$ is the "conformal charge" of the conformal gauge field. Assuming that the connection will be covariant under conformal transformation, the symmetries give:
\begin{equation}
\Gamma^{\lambda}_{\alpha\beta} \rightarrow \Gamma^{\lambda}_{\alpha\beta} \quad , \quad g_{\mu\nu} \rightarrow \Omega(x^\mu)^2 g_{\mu\nu}  \quad , \quad k_{\alpha\beta\gamma} \rightarrow k_{\alpha\beta\gamma}
\end{equation}
\begin{equation*}
A_\mu \rightarrow A_\mu + \frac{2}{e} \partial_\mu \log \Omega(x^\mu)  
\end{equation*}
that the Lagrange multiplier $k_{\alpha\beta\gamma}$ with lower indices does not transform. 
From the variation of the Lagrange multiplier, the condition of Weyl's non-metricity is obtained from the action:
\begin{equation}\label{nmc}
\nabla_\gamma g_{\alpha\beta} = e A_\gamma g_{\alpha\beta}
\end{equation} 
which leads to the solution for the connection:
\begin{equation}
\Gamma^{\rho}_{\mu\nu} = \{ ^\rho_{\mu \nu} \} - \frac{e}{2} g^{\rho\lambda} (g_{\lambda\mu}A_{\nu}+g_{\lambda\nu}A_{\mu}-g_{\mu\nu}A_{\lambda})
\end{equation}
For the conformal invariance we keep quadratic terms of curvatures in the action coupled to the measure $\sqrt{-g}$ which is also conformal invariant in addition to the kinetic term of the gauge fields:
\begin{equation}
\mathcal{L}_{(g,\Gamma)}^{(\textrm{curv})}= \frac{\alpha}{2} R^2 + \frac{\beta}{2} R_{\alpha\beta} R^{\alpha\beta}+\frac{\gamma}{2} R_{\alpha\beta\gamma\delta} R^{\alpha\beta\gamma\delta} 
\end{equation}
\begin{equation}
\mathcal{L}^{(\textrm{Kin})}=-\frac{1}{4} F_{\mu\nu}F^{\mu\nu}
\end{equation}
By introducing non metricity constraint (\ref{nmc}) we can obtain a conformal action and equations of motion, where the basic formulation is the Palatini formalism. The variation with respect to the connection gives the tensors as before:
\begin{equation}
\frac{\delta \tilde{\mathcal{L}}(\kappa)}{\delta \Gamma^{\rho}_{\mu\nu}} = - k^{\mu\beta\nu} g_{\rho\beta} - k^{\nu\beta\mu} g_{\rho\beta}
\end{equation}
with a symmetrization between the components $\mu$ and $\nu$. 
And also the variation with respect to the metric is:
\begin{equation}\label{eomm1}
-\sqrt{-g}G^{\mu\nu}_{(\kappa)} = -\frac{\delta \tilde{\mathcal{L}}(\kappa)}{\delta g_{\mu\nu}} = \tilde{k}^{\mu\nu\lambda}_{;\lambda} + e(\tilde{k}^{\mu\nu\lambda} A_{\lambda} + \tilde{k}^{\nu\mu\lambda} A_{\lambda}) 
\end{equation}
where $\tilde{k}^{\mu\nu\lambda} = \sqrt{-g}k^{\mu\nu\lambda}$. Notice that the conformal change of $\tilde{k}^{\mu\nu\lambda}$ is opposite to conformal charge of $g_{\mu\nu}$, that is transforms as:
\begin{equation}
\tilde{k}^{\mu\nu\lambda} \rightarrow \Omega^{-2} \tilde{k}^{\mu\nu\lambda} 
\end{equation}
Because of the new contribution to the field equation $G^{\mu\nu}_{(\kappa)}$, the complete field equation will contain additional terms which make the field equation to be Weyl invariant, which we will study in details in the future.

If we want to have a linear term in curvature which would save the conformal invariance, we will have to use a modified measure, which is independent of the metric \cite{Guendelman:1996jr}:
\begin{equation}
\mathcal{S} = \int d^4 x \,[ \Phi R + \tilde{\mathcal{L}}_{(\kappa)}+\sqrt{-g} (\mathcal{L}_{(g,\Gamma)}^{(\textrm{curv})} + \mathcal{L}^{(\textrm{Kin})})\,]
\end{equation}
where the construction of this modified measure, for example, is from 4 scalar fields $ \varphi_{a} $, where $ a=1,2,3,4 $. 
\begin{equation}
\Phi=\frac{1}{4!}\varepsilon^{\alpha\beta\gamma\delta}\varepsilon_{abcd}\partial_{\alpha}\varphi^{(a)}\partial_{\beta}\varphi^{(b)}\partial_{\gamma}\varphi^{(c)}\partial_{\delta}\varphi^{(d)}
\end{equation}  
with the symmetries of the scalars and the measure:
\begin{equation}
\varphi_a'\rightarrow\varphi_a'(\varphi) \quad , \quad \Phi' \rightarrow \Phi \, \Omega(x)^2
\end{equation}
with the Jacobian of transformation being $J = \Omega(x)^2$.
This is one option for breaking the metricity condition, using the same Lagrange multiplier.

\section{Discussion}
In this letter we used a Lagrange multiplier in Palatini formalism, which can implement metricity condition, and give the same equation as the field equations which comes from second order formalism. An explicit proof for vanishing of the covariant divergence of the energy-momentum tensor in
beyond Lovelock in Palatini formulation is presented \cite{Koivisto:2005yk}. However by introducing the Lagrange multiplier of the metric, the energy momentum tensor that will appear is the same one in the second order formalism, even the action formulated in the first order. Hence, the stress energy momentum tensor for those theories will be always covariant conserved. A general argument using the path integral approach was formulated as well, and shows the correspondence between the two formalisms, up to the quantum level. 

This mathematical approach discussed in the absence of matter $\mathcal{L}_m$. In the case of $\mathcal{L}_m$ which has no dependence on the connection $\Gamma$, as minimally coupled scalar filed or electromagnetic filed, the calculations are the same. Only in the case of fermion, where we have to use the spin connection formalism, there could be more requirements with different analyses for the correspondence.  

In addition, we used the same method for producing Weyl conformal invariance from an action which is formulated by the first order formalism. The Lagrange multiplier is consistent with conformal invariance and by introducing the proper action a conformal gravity could emerge from this formalism. A complete description of this modified gravity theory will be studied in the future.

The formulation of second order theories in the first order form has a clear advantage from the canonical formulation and therefore concerning the quantization of the theory. The equations of motion are only second order initially, they become higher order when we solve the Lagrange multiplier and reinsert this into the equations of motion. Also the metric has a canonically conjugate momenta, which allows to interpret the integration in the functional integral over $k^{\alpha\beta\gamma}$ and over the metric as an integration in phase space. All these subjects deserve further study.

In the future we will present the physical interpretation of this Lagrange multiplier as the metric conjugate momentum, with the feature of linking between the $1^{st}$ and the $2^{nd}$ order formalism. Effectively, formulating a theory in the $2^{nd}$ order formalism in the form of $1^{st}$ order formalism provides a simple Hamiltonian formulation, because of the fact that your action contains a metric conjugate momentum. This will be used in the context of covariant canonical gauge theory of gravity.  
\acknowledgments
This article is supported by COST Action CA15117 "Cosmology and Astrophysics Network for Theoretical Advances and Training Action" (CANTATA) of the
COST (European Cooperation in Science and Technology). In addition we thank the Foundational Questions Institute FQXi for support, in particular support for our conference BASIC2018, where part of the research was carried out. We thank to Prof. Friedrich W. Hehl for discussions and indites.  


\begin{thebibliography}{0}%
\makeatletter
\providecommand \@ifxundefined [1]{%
 \@ifx{#1\undefined}
}%
\providecommand \@ifnum [1]{%
 \ifnum #1\expandafter \@firstoftwo
 \else \expandafter \@secondoftwo
 \fi
}%
\providecommand \@ifx [1]{%
 \ifx #1\expandafter \@firstoftwo
 \else \expandafter \@secondoftwo
 \fi
}%
\providecommand \natexlab [1]{#1}%
\providecommand \enquote  [1]{``#1''}%
\providecommand \bibnamefont  [1]{#1}%
\providecommand \bibfnamefont [1]{#1}%
\providecommand \citenamefont [1]{#1}%
\providecommand \href@noop [0]{\@secondoftwo}%
\providecommand \href [0]{\begingroup \@sanitize@url \@href}%
\providecommand \@href[1]{\@@startlink{#1}\@@href}%
\providecommand \@@href[1]{\endgroup#1\@@endlink}%
\providecommand \@sanitize@url [0]{\catcode `\\12\catcode `\$12\catcode
  `\&12\catcode `\#12\catcode `\^12\catcode `\_12\catcode `\%12\relax}%
\providecommand \@@startlink[1]{}%
\providecommand \@@endlink[0]{}%
\providecommand \url  [0]{\begingroup\@sanitize@url \@url }%
\providecommand \@url [1]{\endgroup\@href {#1}{\urlprefix }}%
\providecommand \urlprefix  [0]{URL }%
\providecommand \Eprint [0]{\href }%
\providecommand \doibase [0]{http://dx.doi.org/}%
\providecommand \selectlanguage [0]{\@gobble}%
\providecommand \bibinfo  [0]{\@secondoftwo}%
\providecommand \bibfield  [0]{\@secondoftwo}%
\providecommand \translation [1]{[#1]}%
\providecommand \BibitemOpen [0]{}%
\providecommand \bibitemStop [0]{}%
\providecommand \bibitemNoStop [0]{.\EOS\space}%
\providecommand \EOS [0]{\spacefactor3000\relax}%
\providecommand \BibitemShut  [1]{\csname bibitem#1\endcsname}%
\let\auto@bib@innerbib\@empty
\end{thebibliography}%


\begin{thebibliography}{99}
 \bibitem{Palatini} 
 Palatini, A. (1919). "Deduzione invariantiva delle equazioni gravitazionali dal principio di Hamilton". Rend. Circ. Mat. Palermo. 43: 203-212.
 \bibitem{Lovelock} 
Lovelock, D. (1971). "The Einstein tensor and its generalizations". Journal of Mathematical Physics. 12 (3): 498-502. Doi:10.1063/1.1665613. 
\bibitem{Exirifard:2007da} 
  Q.~Exirifard and M.~M.~Sheikh-Jabbari,
  Phys.\ Lett.\ B {\bf 661}, 158 (2008)
  doi:10.1016/j.physletb.2008.02.012
  [arXiv:0705.1879 [hep-th]].
  \bibitem{Gorbunov:2013dqa} 
  D.~Gorbunov and A.~Tokareva,
  Phys.\ Lett.\ B {\bf 739}, 50 (2014)
  doi:10.1016/j.physletb.2014.10.036
  [arXiv:1307.5298 [astro-ph.CO]].
  \bibitem{Myrzakulov:2014hca} 
  R.~Myrzakulov, S.~Odintsov and L.~Sebastiani,
  Phys.\ Rev.\ D {\bf 91}, no. 8, 083529 (2015)
  doi:10.1103/PhysRevD.91.083529
  [arXiv:1412.1073 [gr-qc]].
  \bibitem{Bamba:2014jia} 
  K.~Bamba, R.~Myrzakulov, S.~D.~Odintsov and L.~Sebastiani,
  Phys.\ Rev.\ D {\bf 90}, no. 4, 043505 (2014)
  doi:10.1103/PhysRevD.90.043505
  [arXiv:1403.6649 [hep-th]].
  \bibitem{Sebastiani:2013eqa} 
  L.~Sebastiani, G.~Cognola, R.~Myrzakulov, S.~D.~Odintsov and S.~Zerbini,
  Phys.\ Rev.\ D {\bf 89}, no. 2, 023518 (2014)
  doi:10.1103/PhysRevD.89.023518
  [arXiv:1311.0744 [gr-qc]].
  \bibitem{DeLaurentis:2015fea} 
  M.~De Laurentis, M.~Paolella and S.~Capozziello,
  Phys.\ Rev.\ D {\bf 91}, no. 8, 083531 (2015)
  doi:10.1103/PhysRevD.91.083531
  [arXiv:1503.04659 [gr-qc]].
  \bibitem{Sotiriou:2008rp} 
  T.~P.~Sotiriou and V.~Faraoni,
  Rev.\ Mod.\ Phys.\  {\bf 82}, 451 (2010)
  doi:10.1103/RevModPhys.82.451
  [arXiv:0805.1726 [gr-qc]].
\bibitem{Struckmeier:2017vkf} 
J.~Struckmeier, J.~Muench, D.~Vasak, J.~Kirsch, M.~Hanauske and H.~Stoecker,
  Phys.\ Rev.\ D {\bf 95}, no. 12, 124048 (2017)
  doi:10.1103/PhysRevD.95.124048
  [arXiv:1704.07246 [gr-qc]].
  \bibitem{Struckmeier:2017stl} 
  J.~Struckmeier, P.~Liebrich, J.~Muench, M.~Hanauske, J.~Kirsch, D.~Vasak, L.~Satarov and H.~Stoecker,
  arXiv:1711.10333 [gr-qc].
     \bibitem{Hehl1} 
Hehl, F.W. and Kerlick, G.D. Gen Relat Gravit (1978) 9: 691. $https://doi.org/10.1007/BF00760141$
    \bibitem{Hehl:1976my} 
  F.~W.~Hehl, G.~D.~Kerlick and P.~Von Der Heyde, "Metric-affine variational principles in general relativity. I. Riemannian space-time",
   Phys.\ Lett.\  {\bf 63B}, 446 (1976).
  doi:10.1016/0370-2693(76)90393-2
  \bibitem{Hehl:1981ed}   F.~W.~Hehl, E.~A.~Lord and L.~L.~Smalley,
  Print-81-0283 (COLOGNE).
  \bibitem{Borunda:2008kf} 
  M.~Borunda, B.~Janssen and M.~Bastero-Gil,
  JCAP {\bf 0811}, 008 (2008)
  doi:10.1088/1475-7516/2008/11/008
  [arXiv:0804.4440 [hep-th]].
\bibitem{Olmo:2013lta} 
  G.~J.~Olmo, D.~Rubiera-Garcia,
  Phys.\ Rev.\ D {\bf 88}, 084030 (2013) doi:10.1103/PhysRevD.88.084030
[arXiv:1306.4210 [hep-th]].
\bibitem{Barrientos:2018cnx} 
  E.~Barrientos, F.~S.~N.~Lobo, S.~Mendoza, G.~J.~Olmo and D.~Rubiera-Garcia,
  Phys.\ Rev.\ D {\bf 97}, 104041 (2018)
  doi:10.1103/PhysRevD.97.104041
  [arXiv:1803.05525 [gr-qc]].
 \bibitem{pi}
 Quantization of gauge systems
M. Henneaux, C. Teitelboim. 1992.
Published in Princeton, USA: Univ. Pr. (1992) 520 p
\bibitem{weyl}
   H. Weyl, "Gravitation und Elektrizitat," Sitzungsber. Preuss. Akad. Berlin. 465 - 480
(1918)
  \bibitem{W}
  H. Weyl, Sitzungesber Deutsch. Akad. Wiss. Berli 465 (1918); H. Weyl, Space, Time,
Matter (Dover, New York, 1952).
  \bibitem{Romero:2012hs} 
  C.~Romero, J.~B.~Fonseca-Neto and M.~L.~Pucheu,
  Class.\ Quant.\ Grav.\  {\bf 29}, 155015 (2012)
  doi:10.1088/0264-9381/29/15/155015
  [arXiv:1201.1469 [gr-qc]].
  \bibitem{Guendelman:1996jr} 
  E.~I.~Guendelman and A.~B.~Kaganovich,
  Phys.\ Rev.\ D {\bf 55}, 5970 (1997)
  doi:10.1103/PhysRevD.55.5970
  [gr-qc/9611046].
  \bibitem{Koivisto:2005yk} 
  T.~Koivisto,
  Class.\ Quant.\ Grav.\  {\bf 23}, 4289 (2006)
  doi:10.1088/0264-9381/23/12/N01
  [gr-qc/0505128].
\end{thebibliography}
\end{document}